\documentclass[runningheads]{llncs}

\usepackage{graphicx}
\usepackage{amsfonts}
\usepackage{amsmath}
\usepackage{amssymb}
\usepackage{enumerate}
\usepackage{bm}
\usepackage{color}
\usepackage{wrapfig}
\usepackage{multirow}

\usepackage{hyperref}

\hypersetup{
    colorlinks=true,
    linkcolor=blue,
    filecolor=magenta,      
    urlcolor=blue
    }

\newcommand{\rb}[1]{\left( #1 \right)}
\newcommand{\sqb}[1]{\left[ #1 \right]}
\newcommand{\cb}[1]{\left\lbrace #1 \right\rbrace}

\begin{document}

\title{Constrained Neural Networks for Interpretable Heuristic Creation to Optimise Computer Algebra Systems}

\titlerunning{Constrained Neural Networks for Interpretable Heuristics to Optimise CASs}

\author{Dorian Florescu\inst{1} \and Matthew England\inst{2}}
\authorrunning{D. Florescu and M. England}

\institute{
University of Bath, UK
\and
Coventry University, UK
\email{\\ dmf36@bath.ac.uk}, \quad \email{Matthew.England@coventry.ac.uk}
}

\maketitle

\begin{abstract}
We present a new methodology for utilising machine learning technology in symbolic computation research.  We explain how a well known human-designed heuristic to make the choice of variable ordering in cylindrical algebraic decomposition may be represented as a constrained neural network.  This allows us to then use machine learning methods to further optimise the heuristic, leading to new networks of similar size, representing  new heuristics of similar complexity as the original human-designed one.  We present this as a form of ante-hoc explainability for use in computer algebra development.



\end{abstract}

\section{Introduction}
\label{sec:Intro}

\subsection{Machine learning within computer algebra systems}
\label{subsec:Background}

\emph{Machine Learning} (ML) refers to tools and techniques that learn rules from data, thus allowing a system to improve its performance on a task without any change to the explicit programming.  
ML underpins recent AI advances and is applied in an increasing number of domains.
Mathematics is no exception: ML has been employed to directly perform mathematical computation, such as \cite{LC20} who used ML to integrate expressions and solve ODEs, \cite{BHMRT23} who used ML to find the real discriminant locus, and \cite{He2022} who surveys ML to predict properties from mathematical structures such as groups and graphs.  However, it has been observed 
that mathematical reasoning is an area ML finds difficult. 

Computer Algebra Systems (CASs) are not an obvious domain for ML: their unique selling point is that their answers are exactly correct and so developers are unlikely to replace symbolic computation algorithms with ML\footnote{Experiments like \cite{LC20} conflate two very different causes of failure:  timeout and giving the wrong answer.  For a CAS: the former would be a shame; the latter a disaster.}.  
However, CAS algorithms often come with choices that have no effect on the mathematical correctness of the end result but can have a big impact on the resources required to obtain it, and on how it is presented.  Such choices are often made by human designed heuristics or ``\emph{magic constants}'' \cite{Carette2004} (sometimes not scientifically validated or even documented) but may be better made by ML.

Examples in the literature include \cite{KUV15} which used a Monte-Carlo tree search to find the representation of polynomials most efficient to evaluate, and \cite{SYK16} which used ML classifiers to pick from algorithms that compute the resultant, and \cite{HEDP16} which used ML to decide whether to precondition input for a CAS.

\subsection{Gaining additional mathematical insight from machine learning}
\label{subsec:XAI}

It seems clear that ML can offer optimisation to CASs, but can it offer any further insight into the underlying mathematics / algorithms?
\begin{itemize}
\item \cite{DVBBZTTBBJLWHK21} suggested that ML can help pure mathematicians with the development of new theorems by uncovering patterns in the data.  This led to new results in knot theory and representation theory.
 
\item \cite{RBNBKDREWFKF23} described a form a genetic programming where algorithms were \emph{evolved} with a large language model performing the crossover step.  This resulted in new state-of-the-art heuristics for two NP-hard problems.

\item \cite{PSH20} trained ML to select the next $S$-pair in Buchberger's algorithm to build a Gr\"{o}bner Basis.  An analysis of the agent showed a preference for pairs whose $S$-polynomials are monomials or low degree: prior human-designed heuristics for the problem considered only the $S$-pairs and not the polynomials themselves, so this represents a novel strategy.

\item \cite{PdREC24} described how the SHAP tool \cite{LL17} may be used to analyse ML models that make a heuristic choice for a CAS and then inform \emph{human-level} heuristics --- heuristics that can be expressed in natural language in a similar amount of text as a heuristic designed by a human --- that can be operated without any ML architecture.
\end{itemize}
The tool used in the final paper is from the growing field of Explainable AI.

\subsection{Explainable AI}

Explainable AI (often abbreviated to XAI) may be defined as those ML techniques whose decisions can be explained (at least partially).  Work in the field is usually motivated by the need to error check ML decisions, and to generate greater user trust in ML.  However, in the case of mathematics we hypothesise that XAI tools may be used to give  give guidance or new understanding.  

XAI is a new field: there have been several attempts to give a taxonomy such as \cite{BDDBTBGGMBCH20}.  One distinction in XAI methods that has firmly emerged \cite{Speith2022} is between:
\begin{description}
\item[ante-hoc explainability] which refers to ML methods that are themselves by-design transparent in their decisions; and
\item[post-hoc explainability] which use a secondary analysis of an opaque (i.e. black-box) ML model to generate explanations for it.
\end{description} 
The need for the latter is driven by the so called \emph{performance-explainability trade-of} whereby those ML techniques which allow ante-hoc explainability are thought to give lower accuracy in general.  Although often presented as fact, this trade-off is disputed \cite{Rudin2019} and is likely application dependent \cite{HHWJ22}.  We should also remember that explainability is inherently-audience dependent (compare explainable to an  expert with explainable to the general public) \cite{BDDBTBGGMBCH20}.

SHAP is an example of the post-hoc explainability, forming its explanation through experiments involving perturbations in the input. In the present paper, we consider an alternative approach to the same application as \cite{PdREC24}, but aiming for an ante-hoc explainability approach.

\subsection{Plan of the paper}
\label{subsec:plan}

The paper continues in \S\ref{sec:CAD} with a brief introduction to the CAS choice that we study. Then in \S\ref{sec:FeatGen} we recap a process presented in \cite{FE19} to represent instances of our problem as feature vectors: suitable for use in ML. Our new contributions then follow in \S\ref{sec:NeuralBrown}, where we interpret a well known human-designed heuristic as a small neural network, and in \S\ref{sec:NBSearch} where we search through a family of similar networks to identify an improved human-level heuristic, in effect defining a new type of ante-hoc explainability technique for such problems.

\section{Our application: variable ordering choice for CAD}
\label{sec:CAD}

\subsection{Cylindrical Algebraic Decomposition}
\label{subsec:CAD}

Cylindrical Algebraic Decomposition (CAD) was proposed in \cite{Collins1975} as a method to perform real quantifier elimination.  Given an input in $n$ ordered variables, CAD will decompose the corresponding real space into cells (connected regions of $\mathbb{R}^n$) arranged cylindrically (the projections of a pair of cells with respect to the variable ordering are equal or disjoint) with each semi-algebraic (described by polynomial constraints).  The input to CAD is a set of polynomials and the CAD guarantees that each polynomials will have invariant sign upon each cell of the decomposition.  
CAD has been applied in many fields 
ranging from robotics \cite{MMCRW12} through biology \cite{BDEEGGHKRSW20} to economics \cite{MDE18}.  However, CAD 
but has worst-case complexity doubly exponential in the number of variables \cite{DH88} and thus any work to optimise its implementation can bring swathes of new applications into scope. 

\subsection{CAD variable ordering}
\label{subsec:varord}

The CAD variable ordering controls both the algorithm flow and output format (defining the cylindrical structure).  It can have a huge impact, both practically \cite{DSS04} and in terms of theoretical complexity  \cite{BD07}.  
There exist human-designed heuristics to choose the ordering, e.g. 
\cite{Brown2004}, and 
\cite{dRE22b} which use simple statistics of the input.  
Other heuristics perform increasing amounts of algebraic computation \cite{DSS04}, \cite{BDEW13}, \cite{WEBD14} which we do not consider here in preference for a cheap heuristic.

In the last decade ML models have also been trained to select the variable ordering for CAD.  The first attempt was made by \cite{HEWDPB14} with a support vector machine.  Later, the present authors experimented with a wider range of models \cite{EF19}, methods for feature engineering \cite{FE19} and improved metrics for hyper-parameter selection \cite{FE20a}, culminating in a machine learning pipeline available to use for the task \cite{FE20b}. 
Separately \cite{CZC20} experimented with deep learning for variable selection.

\section{Feature Generation Process}
\label{sec:FeatGen}

A challenge when using ML to optimise a CAS is the communication between them:  the former uses symbolic expressions, and the latter vectors of numerical data called \emph{features}. We summarise next the feature generation process of \cite{FE19}.

\subsection{Formalising Brown's Heuristic}
\label{subsec:Brown}

Our work was based on an analysis of the \texttt{Brown} heuristic \cite{Brown2004} which uses metrics: 
\begin{enumerate}
\item the overall degree in the input of a variable $v$;
\item the maximum total degree of monomials in which variable $v$ occurs; and
\item the number of terms which contain the variable $v$. 
\end{enumerate}
It orders on the earlier metrics, breaking ties with the subsequent ones.

In the following we use index $p$ to refer to polynomials in a problem instance, and index $m$ to refer to the monomials in such a polynomial.  We consider a CAD problem instance as a set of polynomials: 
$\boldsymbol{Pr}=\{\mathcal{P}_p \, \vert \, p=1,\dots,P\}$.
A generic polynomial $P_p$ is then given by a sum of monomials, 
$
\mathcal{P}_p=\sum_{m=1}^{M_p} c^{m,p}\cdot \prod_{i=1}^{n} x_i^{d_i^{m,p}}
$
where $d_{i}^{m,p}$ is the degree of variable $x_i$ in monomial $m$ of polynomial $p$.

Thus the polynomials are defined by the series $\sqb{c^{m,p},(d_1^{m,p},d_2^{m,p},d_3^{m,p})}$, for 
$m=1,\dots,M_p$. This allows formalising the problem set of all polynomials as
\begin{equation*}
\mathbb{S}_{\boldsymbol{Pr}}
= \big\lbrace \big\lbrace \left[ 
c^{m,p},(d_1^{m,p},d_2^{m,p},d_3^{m,p}) 
\right] \, \vert \,  m=1,\dots,M_p
\big\rbrace \, \vert \, p=1,\dots,P \big\rbrace.
\end{equation*}
Then the three metrics of the \texttt{Brown} heuristic above are formalised as
\begin{enumerate}
\item $F^1(d_v) := \max_{m,p} d_v^{m,p}$,
\item $F^2(d_v) := \max_{m,p} \text{sgn}(d_v^{m,p})\cdot(d_1^{m,p}+d_2^{m,p}+d_3^{m,p})$,
\item $F^3(d_v) := \sum_{m,p} \text{sgn}(d_v^{m,p})$.
\end{enumerate}
Here $\max$ and $\sum$ are the maximum and sum functions with the subscript $m,p$ indicating that they are applied over all monomials in all polynomials; while 
sgn($x$) is the function which takes values in $\{-1,0,1\}$ according to the sign of its input (used to identify which terms contain the input $-$ in our situation the sign is only ever positive or zero).

\subsection{Generating similar features}
\label{subsec:FE19}

We notice that the features used by the \texttt{Brown} heuristic are simple to compute using only $\max$, $\sum$, sgn applied to degrees over monomials and polynomials.  We define variants of the functions $\mathrm{max}_x$ and $\sum_x$ where $x$ indicates whether we sum over monomials, polynomials or both; and we define similarly the averaging functions:
\begin{align*}
\text{av}_m &= \frac{1}{M_p}\sum_{m}, \quad
\text{av}_p = \frac{1}{P}\sum_{p}, \quad
\text{av}_{m,p} = \frac{1}{P}\sum_{p}\frac{1}{M_p}\sum_{m};
\end{align*}
then we can express all the features used in \cite{HEWDPB14} in this formalisation \cite{FE19}.  

In\cite{FE19} we generalised this to create a larger set of features: all those of the form 
$
f\left( \boldsymbol{Pr} \right)=(g_4\circ g_3\circ g_2\circ g_1\circ h^{m,p})\left( \boldsymbol{Pr} \right),
$
where 
$
h^{m,p}\left( \boldsymbol{Pr} \right) \in \big\{ d_v^{m,p}, \, 
\text{sgn} \left(d_v^{m,p}\right) \cdot \left( \textstyle \sum_{v'} d_{v'}^{m,p}\right) \, \vert \, v=1,2,3 \big\}
$
and $g_1, g_2, g_3$, $g_4$ are taken from 
$
\{ \textstyle \max_p, \max_m, \max_{m,p},$ $\sum_p, \sum_m, \sum_{m,p}, \text{av}_p, \text{av}_m,\text{av}_{m,p}, \text{sgn}, \textrm{Id} \},
$
with Id as the identity function.  

Many of the features generated will be equivalent:  either at a mathematical level or for the dataset in question and so before using these for ML we should identify a unique subset.  In \cite{FE19} it was demonstrated that we may use these in a ML pipeline to improve the performance compared to using just a human crafted set.  These features have also been the basis for further work in \cite{PdREC24}, \cite{dRE24}.

\section{Interpreting Brown's Heuristic as a Neural Network}
\label{sec:NeuralBrown}

\begin{figure}
    \centering
    \includegraphics[width=0.65\textwidth]{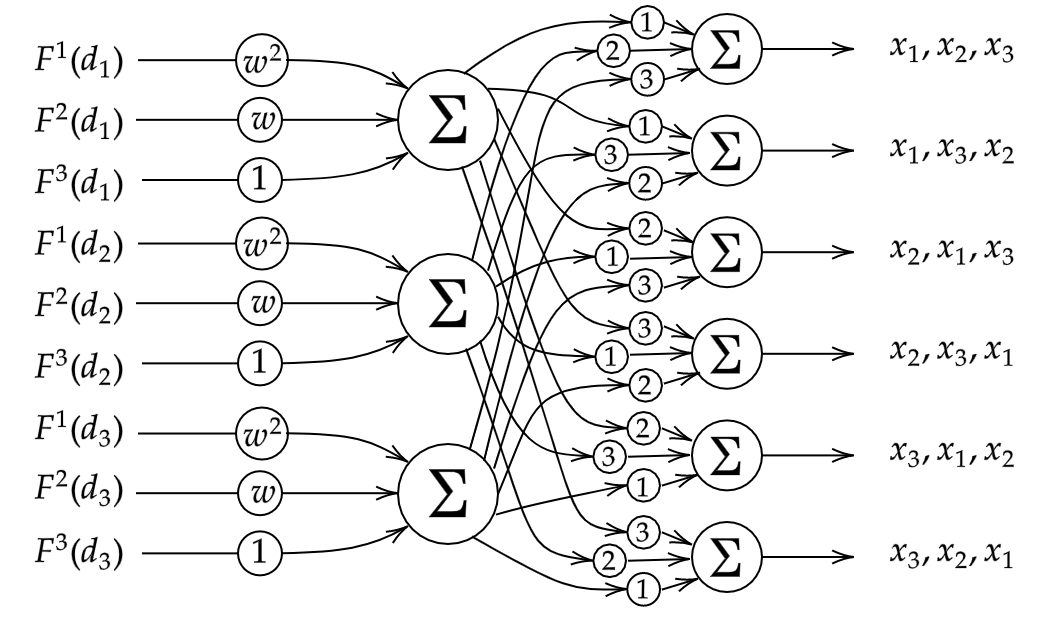}
    \caption{Neural network inspired by Brown's heuristic}
    \label{fig:BrownNN}
\end{figure}

Recall the \texttt{Brown} heuristic from Section \ref{subsec:Brown} which used $3$ rules with different priority levels.
We claim this can be equivalently represented as a dense $2$-layer neural network with summation activation functions, as visualised in Figure \ref{fig:BrownNN} for the three variable case (in the general case there will be $n$ inputs into each of the nodes of the first layer).  The summations are weighted as in Figure \ref{fig:BrownNN} where the weights are defined in terms $w>0$ which we select such that
\begin{equation}\label{eq:w}
    F^i(d_v)<w-1, i,v \in \{1,2,3\}
\end{equation}
for all problems in the dataset (or large enough to cover all problems of interest).

\noindent The outputs of the first network layer are 
\begin{equation*}
    y_v=F^1(d_v)w^2+F^2(d_v)w+F^3(d_v).
\end{equation*}
We will show that the magnitude of $y_v$ orders the variables as Brown's heuristic.

Assume first that $F^1(d_v)>F^1(d_{v'})$, for $v,v'\in\cb{1,2,3}$.  We aim to show $y_v>y_{v'}$ irrespective of the values of the other features. All features are positive integers, meaning our assumption becomes $F^1(d_v)\geq F^1(d_{v'})+1$. 
We start with 
\begin{align*}
    y_{v'}  &= F^1(d_{v'})w^2 + F^2(d_{v'})w+F^3(d_{v'})\\
            &\quad \leq (F^1(d_{v})-1)w^2 + F^2(d_{v'})w+F^3(d_{v'})
\end{align*}
where the inequality follows by our assumption.  Then by the repeated use of (\ref{eq:w}) we have the further strict inequality
\begin{align*}
    y_{v'}  &< F^1(d_{v})w^2 - w^2 + (w-1)w + (w-1) \\
            &\quad = F^1(d_{v})w^2 - 1 
            < F^1(d_{v})w^2.
\end{align*}
Then since all features are positive we have that 
\begin{align*}
    y_{v'}  &< F^1(d_v)w^2 + F^2(d_v)w + F^3(d_v) 
            = y_{v}.
\end{align*}
Now consider the case where $F^1(d_v)=F^1(d_{v'})$ for all $v,v'\in\cb{1,2,3}$ and $F^2(d_v)>F^2(d_{v'})$.  We want to show that under these assumptions $y_{v'}<y_v$ for any realisations of $F^3$.  
We proceed similarly to above: 
\begin{align*}
    y_{v'}  &= F^1(d_{v'})w^2+F^2(d_{v'})w+F^3(d_{v'}) \\
            &= F^1(d_v)w^2 + F^2(d_{v'})w+F^3(d_{v'}) \\
            &\leq F^1(d_v)w^2 + (F^2(d_{v})-1)w + F^3(d_{v'}) \\
            &<    F^1(d_v)w^2 + (F^2(d_{v})-1)w + (w-1)\\
            & = F^1(d_v)w^2 + F^2(d_{v})w -1 \\
            &< F^1(d_v)w^2 + F^2(d_{v})w  \\
            &< F^1(d_v)w^2 + F^2(d_{v})w + F^3(d_{v}) 
            = y_{v}.
\end{align*}
Similarly in the final case where $F^1(d_v)=F^1(d_{v'})$, and $F^2(d_v)=F^2(d_{v'})$ for all $v,v'\in\cb{1,2,3}$ and feature $F^3$ selects the ordering. With $F^3(d_v)>F^3(d_{v'})$
\begin{align*}
    y_{v'}  &= F^1(d_{v'})w^2+F^2(d_{v'})w+F^3(d_{v'})\\
    &= F^1(d_v)w^2 + F^2(d_{v})w+F^3(d_{v'}) \\
    &\leq F^1(d_v)w^2 + F^2(d_{v})w + F^3(d_{v})-1\\ 
    &<    F^1(d_v)w^2 + F^2(d_{v})w + F^3(d_v) 
            = y_{v}.
\end{align*}
Thus the internal layer will be ordered correctly.  The last layer of the network then performs weighted summations of the outputs of the first network layer, $y_v + 2y_{v'} + 3y_{v''}$, with each output neuron labelled corresponding to the weights applied to the variables (see Figure \ref{fig:BrownNN}). If $y_v<y_{v'}<y_{v''}$ then the weighted sum $y_v+2y_{v'}+3y_{v''}$ is the highest among all output neurons, meaning this neural network may be used to produce the same orderings as the \texttt{Brown} heuristic. 

\section{Searching through similar constrained neural networks}
\label{sec:NBSearch}

Now we have represented the \texttt{Brown} heuristic as a (severely constrained) neural network we may consider editing this network to see if a superior heuristic of the same complexity as \texttt{Brown} can be obtained.  


\subsection{Feature Selection}
\label{subsec:NBfeatures}

We perform feature selection using a dataset of random 3-variable polynomial problems (as described in \cite{FE20b}).  We generated $84$ features algorithmically following Section \ref{sec:FeatGen} and taking all permutations this led to $19,656$ possible feature triplets. For each triplet, we computed the variable ordering predicted by the neural network. The triplet that led to the shortest computing time is given by 
\begin{enumerate}
    \item $\sum_{p} \max_{m} d_v^{m,p}$, 
		sum of the highest degree of a variable in each polynomial;
    \item $\sum_{p} \max_{m} \mathrm{sgn}\rb{d_v^{m,p}} \cdot\rb{\sum_{v'} d_{v'}^{m,p}}$, 
    	the maximum sum of degrees of all variables for the terms in which a given variable exists; and
    \item $\sum_{p} \max_{m} \mathrm{sgn} \rb{d_v^{m,p}}$, 
    	the number of polynomials containing the variable.
\end{enumerate}
We note that none of these features are in the \texttt{Brown} heuristic. 

This new feature triplet was selected based on performance on random polynomials.  But to judge its performance we will evaluate it on the NLSAT dataset of 3-variable polynomials from real world (non-linear arithmetic satisfiability problems) as described in \cite{EF19}.  The \texttt{Brown} heuristic requires a total computing time of $10,580\ s$ on this dataset, while the network defined by this new triple resulted in a smaller computing time of $10,181\ s$: which is $399\ s$ shorter.


\subsection{Weight Tuning}
\label{subsec:NBweights}

We next consider changing the weights in the neural network.  This will have the effect of a more complicated combination of the three features to be considered:  but still weighted sums of the same three pieces of information.  

We use the neural network inspired by Brown's heuristic (with $w=30$) as a starting point.  
The weights are then trained on the 3-variable random training dataset using the \emph{adam} stochastic gradient-based optimizer with learning rate $2\cdot 10^{-5}$. From the input data we used the new three features identified in Subsection \ref{subsec:NBfeatures}.  To avoid overfitting, the performance is evaluated for each step on the independent 3-variable NLSAT dataset. 

After only three epochs of training the CAD times decreased to $9908$ (after which there was only minimal improvement).  The weight matrix was only changed slightly, but this was sufficient to avoid the case of ties on the three metrics (followed by a random choice) which was common in the dataset.  

\section{Final Thoughts}
\label{sect:End}

This paper contributes to the ongoing conversation on how ML can contribute to computer algebra: not only algorithm optimisation but also mathematical discovery.  We present an approach, constrained neural networks, that may be viewed as ante-hoc explainability.  It allows for heuristics to be uncovered which are human-level in complexity.  The methodology could be directly applied to other variable ordering choices in symbolic computation, and we expect it could be adapted for use on other choices also.  It remains to be shown whether these more interpretable ML outputs can lead to new mathematical understanding.

\subsection*{Acknowledgements}  

DF and ME were both supported by EPSRC grant EP/R019622/1: \textit{Embedding Machine Learning within Quantifier Elimination Procedures}.  ME was also supported by EPSRC grant EP/T015748/1: \textit{Pushing Back the Doubly-Exponential Wall of Cylindrical Algebraic Decomposition (DEWCAD)}.

%
%
%
%


\end{document}